\RequirePackage{fix-cm}
\documentclass[smallextended]{svjour3}       
\smartqed  
\usepackage{graphicx}
\usepackage{braket}
\usepackage{amsmath}
\usepackage{multirow,bigdelim}
\begin{document}

\title{A thermal quantum classifier}


\author{Ufuk Korkmaz\and Deniz T\"{u}rkpen\c{c}e\and Tahir~\c{C}etin~Ak{\i}nc{\i}  \and Serhat~\c{S}eker}


\institute{
Ufuk Korkmaz \\
\email{ufukkorkmaz@itu.edu.tr} \\
           \and
           Deniz T\"{u}rkpen\c{c}e \\
           \email{dturkpence@itu.edu.tr} \\
			\and			
			Tahir~\c{C}etin~Ak{\i}nc{\i} \\
           \email{akincitc@itu.edu.tr} \\
			\and
			Serhat~\c{S}eker \\
           \email{sekers@itu.edu.tr} \\
			\at Department of Electrical Engineering, \.{I}stanbul Technical University, \.{I}stanbul, 34469 Turkey
}

\date{Received: date / Accepted: date}

\maketitle

\begin{abstract}
A data classifier is the basic structural unit of an artificial neural network. These classifiers, known as perceptron, make an output prediction over the linear summation of the input information. Quantum versions of artificial neural networks are considered to provide more efficient and faster artificial intelligence and learning algorithms. The most generic and realistic open quantum systems are the quantum systems in thermal environments and the information carried by the thermal reservoirs is the temperature information. This study shows that an open quantum system that is in contact with many information channels is a natural information classifier. More specifically, it has been demonstrated that a two-level quantum system can classify temperature information of distinct thermal reservoirs. The results of the manuscript are of importance to the construction of thermal quantum neural networks and the development of minimal quantum thermal machines. Also, a physical model, proposed and discussed with realistic parameters, shows that faster operating thermal quantum classifiers can be built than the classical versions.  
\end{abstract}

\section{Introduction}

Artificial neural networks mimic the learning models inspired from the biological context and find a vast variety of applications on data processing~\cite{mcculloch_logical_1943,schmidhuber_deep_2015,ngai_application_2009,misra_artificial_2010,Hou_2009,Tajbakhsh_2016,Tang_extreme_2016,Shi_End_2017,Halil_Energy_2019,gu_recent_2018,Halil_cornell_2019}. Quantum versions of neural nets~\cite{rebentrost_quantum_2014,schuld_simulating_2015,banchi_quantum_2016,schuld_quantum_2018,yamamoto_simulation_2018} are expected to have speed or resource superiorities against their classical versions through the non-classical quantum resources such as quantum entanglement or quantum coherence~\cite{horodecki_quantum_2009,streltsov_2017,turkpence_decoherence_2018,turkpence_quantum_2016,turkpence_photonic_2017}. 
To this end, several algorithmic models of simple neural nets have been proposed to be implemented on a universal quantum computer where the present computational advantages of quantum computing have been utilized~\cite{tacchino_artificial_2019,wan_quantum_2017,torrontegui_unitary_2019,riste_demonstration_2017}. Instead, quantum systems could also be simulated through analog quantum simulation~\cite{georgescu_quantum_2014} where the physical properties, e.g., speed of the given task becomes significant.   

In this study, we introduce an experimentally accessible quantum system operating in a thermal environment as a data classifier. The possibility of the minimal machines~
\cite{man_smallest_2017,linden_how_2010,gelbwaser-klimovsky_minimal_2013,brunner_virtual_2012}
  operating in the nano or micro scale is appealing due to their extremely small dimensions and the possible speedups by their operating frequency range. This motivates us to devise an open quantum perceptron operating orders of magnitude faster than classical ones. A binary classifier or in mathematical terms, a perceptron is the basic unit of an artificial neural network and returns a binary decision modulated by an activation function corresponding to the weighted linear combination of the data inputs. Likewise, the quantum model we introduce returns a binary decision corresponding to the linear combination of the temperature data of the connected thermal reservoirs.

In general any observable of an open quantum system weakly coupled to the reservoir degrees of freedom equilibrates to a steady value in the long-time limit regardless of the system's initial state~\cite{reimann_foundation_2008,linden_quantum_2009,breuer_theory_2007}. An open quantum system experiencing such an evolution is said to be connected to a Markovian reservoir. On the other hand, the system could also experience a non-Markovian reservoir, in which the past states of the system affect future evolution.~\cite{breuer_theory_2007}. We limit the scope of the present study with the Markovian thermal quantum reservoirs with finite temperatures. When the equilibrated state of a quantum system reaches a Gibbs state, the system is said to be thermalized. According to the general view of quantum information, thermal noise is a major obstacle to protect the valuable quantum state. To this view, the equilibrated quantum state is a highly mixed state in which the useful quantum information is irreversibly lost. 

However, according to the novel approaches, quantum reservoirs are not necessarily the garbage cans in which the quantum information is thrown, however, they can be referred to as information channels in which the reservoir information transmitted~\cite{blume-kohout_simple_2005,zwolak_redundancy_2017}. In compliance with this approach, the system is said to be thermalized when the quantum system temperature is equal to the thermal reservoir temperature, that is, the temperature information can be considered as `has been sent' to the system by the reservoir. ~\cite{liao_single-particle_2010}. If the system is in contact with multiple thermal reservoirs with different temperatures, an effective temperature can be defined as the thermalized state 
of the system depending on the reservoir temperatures. This effective temperature is referred to as virtual temperature~\cite{brunner_virtual_2012} and plays a significant role in understanding the thermal quantum devices.~\cite{silva_performance_2016}.  

More particularly, we investigate a two-level quantum system in contact with thermal environments, in general, corresponding to different temperatures. In this scheme, the temperatures of the environments are considered as the input data and the corresponding effective temperature of the two-level system in the equilibrated steady state is introduced as the binary decision of the system. We adopt the standard Lindblad formulation
~\cite{lindblad_generators_1976,gorini_completely_1976} for the open quantum system evaluation and obtain the reduced dynamics by tracing out the reservoir degrees of freedom. The main idea of the study relies on the complete positivity (CP), divisibility and additivity of the quantum dynamical maps
~\cite{wolf_dividing_2008,filippov_divisibility_2017,kolodynski_adding_2018}.
Just like a classical perceptron, the proposed quantum system experience the summation of the temperature data from different environments thanks to the weighted convex summation of the quantum dynamical maps~\cite{filippov_divisibility_2017}. We both analytically and numerically demonstrate the response of the system calculating the steady state temperature in terms of the level populations. 

\begin{figure}
\centering
\includegraphics[width=3.5in]{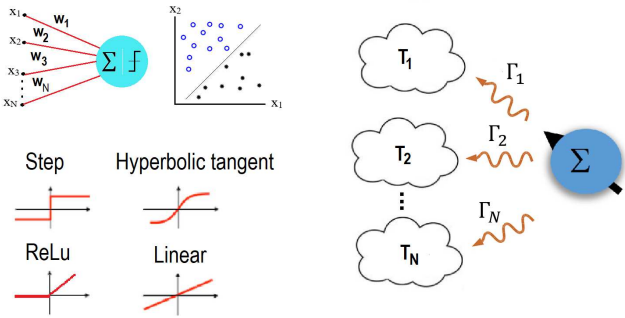}
\caption{Classical (left panel) and quantum (right panel) versions of a perceptron with $N$ data inputs. 
An accurately functioning perceptron linearly separates the data instances (top right of the left panel) with mostly used activation functions (bottom of the left panel). In the open quantum version of the perceptron (right panel), a single spin dissipation through $N$ distinct thermal reservoirs (with relaxation rates $\Gamma_i$) is referred to as temperature information flow to the quantum system (see the text) through $N$ distinct quantum channels.}\label{fig:fig1}
\end{figure}

We demonstrate that the proposed quantum system returns a binary decision  for the temperature data of the thermal reservoirs in the steady state limit depending on the temperature values as well as the decay rates of the system to the reservoirs. We also show that the returned data is linearly separable, that is, the suggested system faithfully classifies the input temperature information. A physical model of the system was proposed and discussed with realistic parameters. 

\section{Framework and system dynamics}\label{SecFrame}

The simplest model for binary data classification can be described by the perceptron (See Fig. 1 upper left panel) in which returns a binary decision corresponding to a weighted summation of various data inputs (features) modulated by specific functions. The input data $x_{1}, x_{2},...,x_{N}$ are composed of any mensurable individuals with their associated weights $w_{1}, w_{2},...,w_{N}$. The perceptron algorithm returns a binary result $f(y)$ that has been modulated by an activation function $f(\cdot)$, corresponding to an input $y=\Sigma_i x_i w_i$. A  few of commonly used activation functions are depicted in the bottom of the left panel of Fig. 1. For example, a step function returns $f(y)=1$ if $y\geq 0$ and returns $f(y)=-1$ else. On the other hand, a linear function yields a linear and continuous response for any $y$. A properly functioning perceptron can linearly separate the data instances corresponding to multi-dimensional feature space. 

As stressed in the previous section, we introduce a quantum version of a data classifier operating in a thermal environment. The right panel of Fig. 1 represents a small quantum system simultaneously interacting with multiple thermal environments characterized by respective temperatures. We adopt a quantum master equation (QME) approach, to describe the open quantum dynamics, as a standard tool~\cite{breuer_theory_2007}.  As a general view, a time local QME can be described as \cite{gorini_completely_1976}
\begin{equation}
\frac{d}{dt}\varrho_s (t)=\mathcal{L}_t[\varrho_s(t)]=-\frac{i}{\hbar}[\mathcal{H}(t),\varrho_s]+\mathcal{D}_t[\varrho_s(t)]
\end{equation}
where $\mathcal{L}_t$ is the dynamical generator which can be decomposed into unitary and dissipative parts. Here, $\varrho_s$ is the density matrix of the system of interest, $\mathcal{H}(t)$ is any Hermitian Hamiltonian representing the system and $\mathcal{D}_t$ stands for the dissipative part of the generator induced by environmental effects. A physically valid QME should lead to a dynamical map $\Lambda_t$ satisfying $\Lambda_t[\varrho_s(0)]=\varrho_s(t)$ for any $t\geq 0$. That is, the map should preserve the density matrix properties such as positivity and trace unity. Moreover, the map should also hold complete positivity for any point of the evolution. The dynamical map ensuring all these properties is said to be completely positive trace preserving dynamical map (CPTP). A CPTP map can exhibit a non-monotonic character with memory effects due to the time local feature of the generator $\mathcal{L}_t$ yielding a non-Markovian evolution. Note that we are only interested in the weak coupling regime in which the open system evolution is Markovian. In this particular form, the Hamiltonian and the dissipative term is time-independent. This form of the generator is said to be in the standard  Lindblad form and, in general, can be represented as

\begin{equation}
\mathcal{L}[\varrho]=-\frac{i}{\hbar}[\mathcal{H},\varrho]+\mathcal{D}[L_{\varrho}]
\end{equation}
where 
\begin{equation}\label{Dissipation_term}
\mathcal{D}[L_{\varrho}]=\sum_k \Gamma_k[L_k\varrho L_k^{\dagger}-\lbrace L_k^{\dagger}L_k,\varrho\rbrace/2].
\end{equation}

Here, $L_k$ is the jump operator acting on each subsystem independently, $\lbrace \cdot \rbrace$ is the anti-commutation and $\Gamma_k$ is the relaxation rate of each subsystem through the reservoir. Note that in this scheme, the dynamics form a semigroup in which the generator and the map is related with an 
$\Lambda_t =\texttt{exp}[t\mathcal{L}]$ exponential form for all $t$. Moreover, the map representing this evolution is called CP divisible since it satisfies the composition law 
$\Lambda_{t_2,t_0}=\Lambda_{t_2,t_1}\Lambda_{t_1,t_0}$ for all $t_2\geq t_1\geq t_0\geq 0$ ~\cite{filippov_divisibility_2017,kolodynski_adding_2018}. 

As mentioned in the previous section, we are interested in the effective temperature of a small quantum system at steady state in the presence of different thermal reservoirs. In this respect, we define the system evolution by exploiting the additivity of the dynamical generators such that
\begin{equation}\label{Eq:ConvexL}
\mathcal{L}[\varrho]=\frac{\partial \varrho}{\partial t}=P_1\mathcal{L}^{(1)}_t+\ldots +P_N\mathcal{L}^{(N)}_t
\end{equation}
where $P_i\neq 0$ are the probabilities of encountering the system  from the $ith$ reservoir obeying the unity condition $\sum_i P_i=1$.  Equation (\ref{Eq:ConvexL}) is a valid physical evolution if and only if each generator provides the CP divisibility and the weak coupling condition to the reservoirs~\cite{filippov_divisibility_2017,kolodynski_adding_2018}. In general, the bath consists of a large number of harmonic oscillators with various modes of bath frequencies $\omega_j$. Therefore the coupling rates to the reservoir $\Gamma_k(\omega_j)$ and the corresponding jump operators $L_k(\omega_j)$ are the functions of bath frequencies in (\ref{Dissipation_term}). However, for simplicity, we will encounter two-level systems with a specific frequency $\omega$ and a corresponding relaxation rate $\Gamma$ as the bath degrees of freedom in Gibbs state denoted by 
\begin{equation}\label{ThermalB}
\varrho_B=\frac{\texttt{exp}[-\beta H_B ]}{\texttt{Tr}_B \texttt{exp}[-\beta H_B]}
\end{equation}
where $H_B$ is the bath Hamiltonian, $\beta=1/k_B T$ is the inverse bath temperature and $k_B$ is the Boltzmann constant. 
Each generator in (\ref{Eq:ConvexL}) can be defined as~\cite{breuer_theory_2007}
\begin{equation}\label{ThermalLindblad}
\mathcal{L}^{(i)}_t=-\frac{i}{\hbar}[\mathcal{H},\varrho]+\Gamma^{(i)}(\bar{n}^{(i)}+1)\mathcal{D}[L_{\varrho}]+\Gamma^{(i)}\bar{n}^{(i)}\mathcal{D}[L^{\dagger}_{\varrho}]
\end{equation} 
with
\begin{equation}\label{Excitation}
\bar{n}^{(i)}=[\texttt{exp}(\frac{\hbar\omega}{k_B T^{(i)}})-1]^{-1}. 
\end{equation}

Here, $\bar{n}^{(i)}$ is the temperature dependent excitation number of $ith$ thermal bath with  corresponding temperature $T^{(i)}$ and $\Gamma^{(i)}$ is the relaxation rate of the system to $ith$ bath and $\hbar$ is the reduced Planck constant. Note that (\ref{Eq:ConvexL}) is the quantum equivalent weighted summation of the perceptron input temperature data. The response of the system returns an effective (or virtual) system temperature $T_S$ in the thermalized state
\begin{equation}
\varrho_S=\frac{\texttt{exp}[-\beta_S H_S ]}{\texttt{Tr}_S \texttt{exp}[-\beta_S H_S]}
\end{equation}
where $\beta_S=1/k_B T_S$. The effective temperature $T_S$ can easily be defined by means of the level populations of the two-level system as ~\cite{quan_quantum_2007,silva_performance_2016}
\begin{equation}
T_S=\frac{\hbar\omega_S}{k_B\texttt{ln}[\frac{p_g}{p_e}]} \label{QbitTemp}
\end{equation}
for the thermalized state of $\varrho_S$ where, respectively, $p_g$ and $p_e$ are the ground and excited state populations. The effective temperature in (\ref{QbitTemp}) is the identifier of the response of the system in the steady state corresponding to various relaxations to the independent thermal reservoirs. 

\section{The quantum classifier}

In this section, we describe the introduced model with  a simple physical example and examine the system dynamics numerically. In the modelled example, a spin-$1/2$ system couples to $N$ thermal reservoirs, in general, with different finite temperatures and different relaxation rates. Equations (\ref{Dissipation_term}) and (\ref{ThermalLindblad}) characterizes the evolution of the system. We define the system Hamiltonian and the jump operators as $\mathcal{H}=\omega_S\sigma_z/2$ and  $L\equiv \sigma^{-}=\ket{g}\bra{e}$, $L^{\dagger}\equiv \sigma^{+}=\ket{e}\bra{g} $ where $\sigma_z=\ket{e}\bra{e}-\ket{g}\bra{g}$, $\sigma^{+}$ and $\sigma^{-}$ are, respectively, the Pauli-$z$, Pauli-raising and -lowering operators. The generalization of (\ref{ThermalLindblad}) to $N$ thermal reservoirs will lead to a microscopic master equation 
\begin{equation}\label{MasterEq}
\dot{\varrho}=-\frac{i}{\hbar}[\mathcal{H},\varrho]+\sum_i^N \Gamma^{(i)}\left[(\bar{n}^{(i)}+1)\mathcal{D}[L_{\varrho}]+\bar{n}^{(i)}\mathcal{D}[L^{\dagger}_{\varrho}]\right]. 
\end{equation}
 Note that (\ref{MasterEq}) obeys (\ref{Eq:ConvexL}) in the weak coupling limit where $P_i\sim \Gamma^{(i)} $. This model which can be illustrated in the right panel of Fig. 1 is very generic and has been studied with various motivations~\cite{fedortchenko_finite-temperature_2014,chan_quantum_2014,shabani_artificial_2016}. 
 
As stated in the previous section, the objective of the current study is to demonstrate that the response of the system in the equilibrated state could serve as a quantum classifier that linearly classifies the temperature data of the connected thermal reservoirs. To this end, we analytically solve (\ref{MasterEq}) and obtain the temperature in terms of relaxation rates at steady state. Inserting the two-level system Hamiltonian and the relevant operators into (\ref{MasterEq}) and by taking $\dot{\varrho}=0$, we obtain (see Appendix) the system populations ratio at the steady state as
\begin{equation}\label{Ratio}
\frac{p_g}{p_e}=\frac{\sum_i(\bar{n}^{(i)}+1)\Gamma^{(i)}}{\sum_i \bar{n}^{(i)}\Gamma^{(i)}}.
\end{equation}
The steady state temperature can be obtained by inserting (\ref{Ratio}) into (\ref{QbitTemp}) without any approximation. On the other hand, using some approximations for the particular case where all the coupling rates are equal $(\Gamma^{(1)}=\Gamma^{(2)}=\cdots = \Gamma^{(N)})$ the steady state temperature $(T_S)^{ss}$ of the system reduces to 
\begin{equation}\label{AvarageT}
(T_S)^{ss}\cong \frac{T_1+\cdots T_N}{N}=\frac{\sum T_i}{N}=\bar{T}_N.
\end{equation}
That is, for this particular case, steady state temperature is the arithmetic mean $\bar{T}_N$
of the bath temperatures. Therefore, we define the binary classification of our thermal classifier as $class_1$ for $(T_S)^{ss}\geq \bar{T}_N$ and as $class_2$ for $(T_S)^{ss} < \bar{T}_N$.

\begin{figure}
\centering
\includegraphics[width=2.8in]{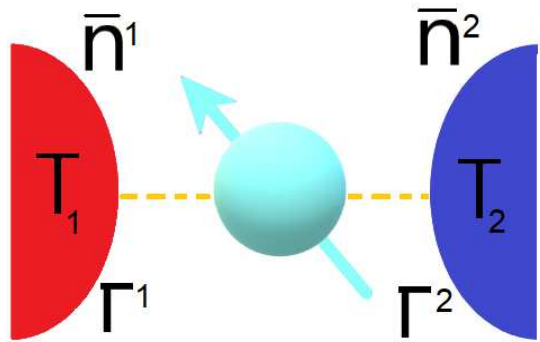}
\caption{ Single-$1/2$ spin system coupled to two thermal reservoirs with $\bar{n}^1$ and $\bar{n}^2$ excitation numbers and $T_1$, $T_2$ finite temperatures. The system relaxes to the reservoirs with the rates $\Gamma^1$ and $\Gamma^2$.}
   \label{fig:fig2}
\end{figure}

\subsection{Numerical Results}

Before demonstrating the classification process, first, we investigate the thermalization dynamics in the time domain.  Note that we choose the relaxation rates of the system to the reservoirs $\Gamma\ll \omega_S$ much smaller than the characteristic system frequency consistent with the weak coupling condition. In the rest of the manuscript, we take $\hbar$,
$k_B=1$ for the calculations. Here, we present a simple example of our model encountering the dynamics of the two-level system relaxing to two thermal baths with different temperatures as depicted in Fig.~\ref{fig:fig2}. The quantum states of these thermal baths are defined by (\ref{ThermalB}) with corresponding temperatures $T_1$ and $T_2$ in contact with the system of interest. We choose the initial state of the system as the ground state $\varrho_S=\ket{g}\bra{g}$ corresponding to zero temperature.

\begin{figure}
\centering
\includegraphics[width=3.4in]{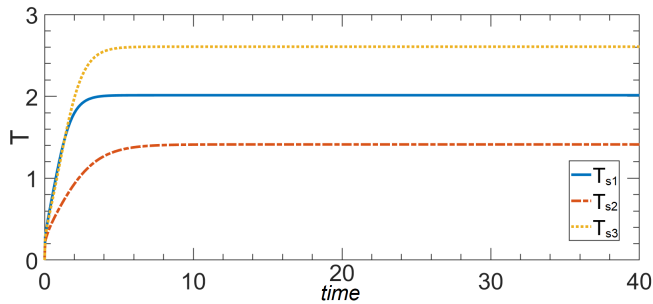}
\caption{ The relaxation dynamics of the system in terms of the system temperature depending on different coupling rate pairs. Reservoir temperatures are fixed and $T_1=3$ and $T_2=1$ in units of $\hbar\omega/k_B$. The reservoir frequency and the system frequency are, respectively, $\omega=\omega_S$ equal. Three cases were considered with three different relaxation rate pairs as $\Gamma^1=\Gamma^2=0.1$ (solid line) ; $\Gamma^1=0.1$, $\Gamma^2=0.05$ (dotted line) and $\Gamma^1=0.05$, $\Gamma^2=0.1$ (dotted-dashed line). Time is dimensionless and scaled by $\omega$.}
   \label{fig:fig3}
\end{figure}

As shown in Fig.~\ref{fig:fig3}, when we solve the microscopic master equation given in (\ref{MasterEq}), we obtain the thermalization curves for different relaxation rate pairs of the system to the baths. The system rapidly reaches the steady state with a definite temperature obtained by (\ref{QbitTemp}). Note that, the thermalized system temperature is between the temperatures of the baths. Moreover, the system temperature approaches to the bath temperature in which the corresponding relaxation rate is larger. If the relaxation rates are $\Gamma^1=\Gamma^2$ equal (solid line of Fig.~\ref{fig:fig3}) the system thermalizes to the average of the baths' temperatures justifying (\ref{AvarageT}). This simple result shows that the additivity of dynamical maps, defined in (\ref{Eq:ConvexL}), work well in this thermal quantum state example with the corresponding parameters. Another crucial point is that the relaxation rates $\Gamma^{(i)}$ play the role of weights $w_i$ as in the classical perceptron example. Therefore, both the bath temperatures and the corresponding relaxation rates are the ingredients of the input data for the introduced quantum classifier. In the classical scheme, the linear separation of a classifier is illustrated in the feature space as well as the weight space. 
Accordingly, next, we present  
the linear separation of the data instances in both the relaxation rate and the temperature space for our the quantum classifier model. But first, we investigate the response of the system, again, in terms of the steady state system temperature under the linear variation of input data parameters $\Gamma^{(i)}$.

\begin{figure}
\centering
\includegraphics[width=3.4in]{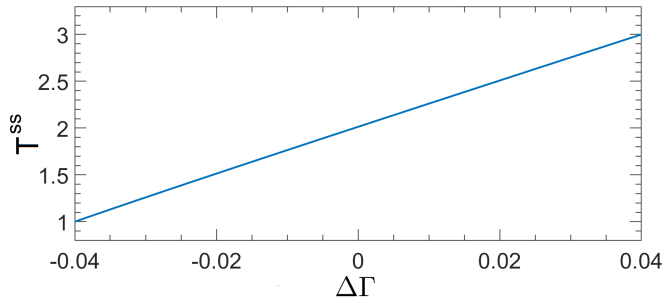}
\caption{The steady response of the system in terms of system temperature depending on coupling rates to the thermal baths. $\Gamma^1=\Gamma/2+\Delta\Gamma$ and $\Gamma^2=\Gamma/2-\Delta\Gamma$ where $\Gamma=0.08$. Reservoir temperatures are fixed and $T_1=3$ and $T_2=1$ in units of $\hbar\omega/k_B$. The reservoir frequency and the system frequency are, respectively, $\omega=\omega_S$ equal.}
   \label{fig:fig4}
\end{figure}

In this part of our example, we define the relaxation rates to the thermal baths as $\Gamma^1$ and $\Gamma^2$ where $\Gamma^1=\Gamma/2+\Delta\Gamma$ and $\Gamma^2=\Gamma/2-\Delta\Gamma$.
Here, $\Delta\Gamma$ ranges between $-\Gamma/2\leq\Delta\Gamma\leq\Gamma/2$. We use the same parameters of Fig.~\ref{fig:fig3} and plot the steady state temperature of the system depending on the values of $\Gamma^1$ and $\Gamma^2$. Note that, for $\Delta\Gamma=\Gamma/2$, $\Gamma^1=\Gamma$ while $\Gamma^2=0$. In this case, the system is coupled only to the first reservoir, therefore, the steady state temperature of the system is equal to the temperature $T_1$ of the first reservoir. The situation is the opposite when $\Delta\Gamma=-\Gamma/2$. As obvious in Fig.~\ref{fig:fig4}, the steady state response of the system is linear against the linear variation of $\Delta\Gamma$. This corresponds to a linear activation function-like behaviour for our quantum classifier. Similar results were reported regarding the thermal environments in the past~\cite{romero_is_2004}. However, a non-linear response character with respect to the linear variation of input parameters was recently reported in case of information reservoirs~\cite{turkpence_steady_2019}. 

\begin{figure}
\centering
\includegraphics[width=3.4in]{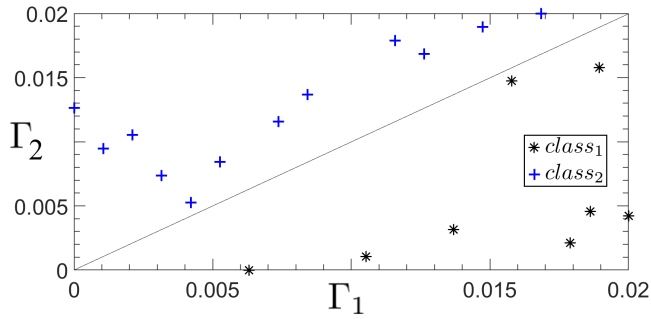}
\caption{ Classification of the temperature data instances by the quantum model in $\Gamma$ space. The temperatures of the baths are $T_1=T_2=3$ equal in units of $\hbar\omega/k_B$. There are randomly selected 20 relaxation rate pairs and the classified instances are linearly separable.}
   \label{fig:fig5}
\end{figure}

\begin{figure}
\centering
\includegraphics[width=3.4in]{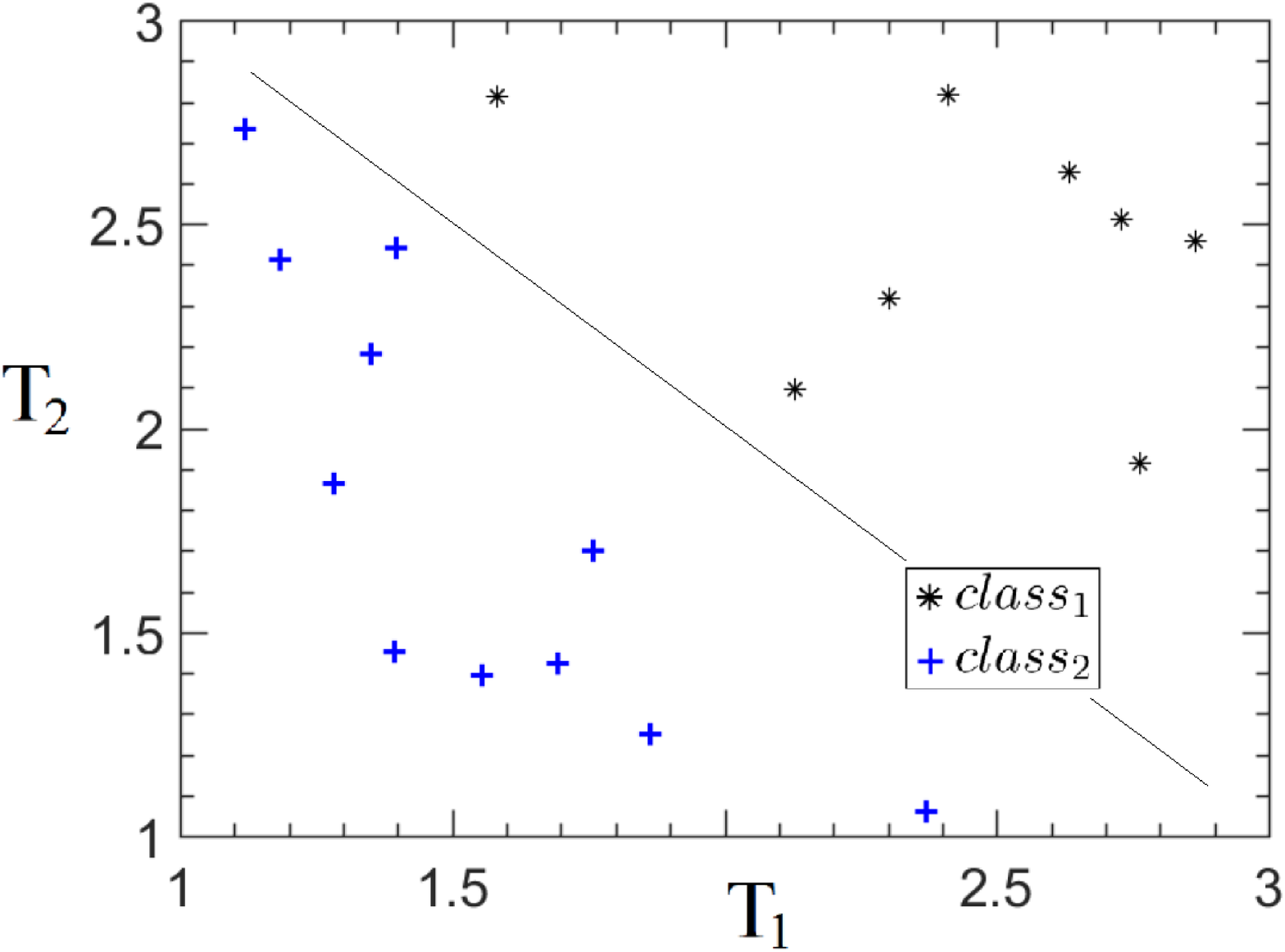}
\caption{ Classification of the temperature data instances by the quantum model in $T$ space.
The relaxation rates of the system to the baths are $\Gamma^1=\Gamma^2=0.02$ equal.
There are randomly selected 20 temperature pairs and the classified instances are linearly separable.}
   \label{fig:fig6}
\end{figure}

After these analyses, we present the success of the linear separation of the data instances of the quantum classifier model. For a clear demonstration, we encounter two cases. First, (see Fig.~\ref{fig:fig5}) the temperatures of the baths are equal and fixed and the separation of the data instances are illustrated in the $\Gamma$ space. Second, the relaxation rates are equal and fixed and the separation of the data instances are illustrated (see Fig.~\ref{fig:fig6}) in the $T$ space.

Note that all the points in Figs.~\ref{fig:fig5} and ~\ref{fig:fig6} are the steady states of the two-level system between the reservoirs. These results show that the introduced quantum model reliably classify the temperature data instances in the thermal environment in which it operates. Though these examples are given for the two thermal reservoirs, the introduced model always returns a binary decision regardless of the number of thermal environments as it obeys (\ref{Eq:ConvexL}).

\subsection{Collision model for open quantum system dynamics}

Before proposing a physical apparatus to represent the quantum classifier, we describe the collisional model to implement the open system dynamics. Collisional models have been recently become popular by their versatile representation schemes of the open quantum systems~\cite{filippov_divisibility_2017,scarani_thermalizing_2002,lorenzo_composite_2017,strasberg_quantum_2017,bruneau_repeated_2014}. 
Note that the validity of the proposed classifier hinges on the Markovianity of the open system dynamics that hold CP divisibility and the weak coupling conditions. Here, we outline the general framework in which the scheme is equivalent to the Markovian dynamics.

\begin{figure}
\centering
\includegraphics[width=3.0in]{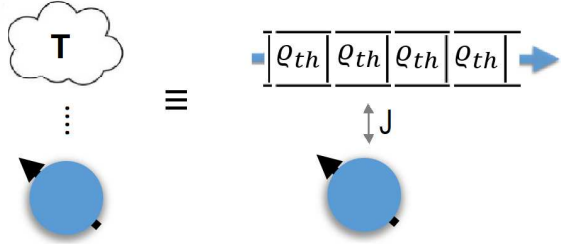}
\caption{ The collision model representing the open quantum system dynamics. A thermal reservoir with temperature $T$ (left) in contact with a single spin system was simulated by discrete repeated interactions (right). }
   \label{fig:fig7}
\end{figure}

As depicted in Fig.~\ref{fig:fig7}, identically prepared thermal units $\lbrace \mathcal{R}_n \rbrace$ (each with the temperature $T$) sequentially interact with the system qubit $\mathcal{S}$ with identical interaction time $\tau$. It's supposed that, initially, system-reservoir state $\mathcal{SR}$ is in a separable $\varrho(0)=\varrho_{\mathcal{S}}(0)\otimes \varrho_{\mathcal{R}}$ state where initial system state is $\varrho_{\mathcal{S}}(0)=\ket{g}\bra{g}$ corresponding to zero temperature. On the other hand, initially, each unit is in a thermal Gibbs state with temperature $T$. According to the standard Markov description of the collision model we follow, the ancillas are discrete, identical and do not interact with each other.  We define the unitary propagator $\mathcal{U}_{\mathcal{SR}_n}=\texttt{exp}[-i\mathcal{H}_{\mathcal{SR}_n}\tau]$ where the reduced Planck constant  was taken $\hbar=1$. Here, $\mathcal{H}_{\mathcal{SR}}$ is the time-independent interaction Hamiltonian denoting no time-local evolution during the ancilla-system evolution. We define this Hamiltonian as 
\begin{equation}\label{Flip}
\mathcal{H}_{\mathcal{SR}_n}=\frac{h}{2}(\sigma_z^n+\sigma_z^s)+J(\sigma_{+}^n\sigma_{-}^s+h.c.).
\end{equation}
where $\sigma_z^n$ and $\sigma_{\pm}^n$ are the Pauli matrices act on the $nth$ reservoir units, $\sigma_{\pm}^s$ are the Pauli matrices act on the system. Here, $J$ is the coupling constant between the system and the single reservoir unit and $h$ is the frequency of the system and the ancilla.  

Each interaction defined above, yields a quantum dynamical map 
\begin{equation}
\Phi_{\mathcal{SR}}[\varrho]=\mathcal{U}_{\mathcal{SR}}\left(\varrho_{\mathcal{SR}}^0\right)\mathcal{U}_{\mathcal{SR}}^{\dagger}
\end{equation}
where $\mathcal{U}_{\mathcal{SR}}$ is the successive implementation of $\mathcal{SR}$. In this Markov scheme, after a sufficient number of interactions, the system state evolves into an identical state that of one of the reservoir units. That is, the system temperatures reach the thermal reservoir temperature by this discrete dynamical evolution. This process is known as quantum homogenization~\cite{scarani_thermalizing_2002}. By `sufficient', we imply that after a sufficient collision number $n$, the system ends up with a steady state as
\begin{align}
\varrho_{\mathcal{S}}^n=&\texttt{Tr}_n   \big[ \mathcal{U}_{\mathcal{SR}_n}\ldots\texttt{Tr}_1[\mathcal{U}_{\mathcal{SR}_1}\left(\varrho_{\mathcal{S}}^0\otimes\varrho_{\mathcal{R}_1}\right)\mathcal{U}_{\mathcal{SR}_1}^{\dagger}]\otimes\ldots \ldots\otimes\varrho_{\mathcal{R}_n}\mathcal{U}_{\mathcal{SR}_n}^{\dagger} \big]
\end{align}
where $\texttt{Tr}_i$ is the partial trace over $ith$ unit. The dynamical maps explained above could also be represented like
\begin{equation}\label{Concatn}
\varrho_{\mathcal{S}}^n=\Lambda_n\circ\Lambda_{n-1}\circ\ldots\circ\Lambda_1\equiv\Lambda_n[\varrho_{\mathcal{S}}^0]
\end{equation}
where $\Lambda_i[\varrho_{\mathcal{S}}]=\texttt{Tr}_i[\mathcal{U}_{\mathcal{SR}_i}\left(\varrho_{\mathcal{S}}\otimes\varrho_{\mathcal{R}_i}\right)\mathcal{U}_{\mathcal{SR}_i}^{\dagger}]$. Here, the operation `$\circ$' is known as concatenation satisfying $\Lambda_2\circ\Lambda_1[\varrho]=\Lambda_2(\Lambda_1[\varrho])$. One should keep in the mind that the dynamical maps expressed above, preserves the CP property as well as trace unity. Moreover, each map corresponding to each collision is a CPTP map. In the physical model, we assume that the coupling to the reservoir units is much smaller than the characteristic frequencies $J\ll h$, that is, the system also satisfies the CP divisibility~\cite{filippov_divisibility_2017}. 
Therefore, the collision model in which we prefer to implement to describe the open system dynamics is consistent with the Markov Lindblad formulation expressed in Section~\ref{SecFrame}. 

As expressed in the introduction, at the end of the open system evaluation the system thermalizes with the bath. Fig.~\ref{fig:fig7} illustrates the coupling of the two-level quantum system to a finite temperature thermal bath by a collision model expressing the evolution by small discrete steps mathematically represented by (\ref{Concatn}). As we are interested in the steady behaviour of the system in the presence of $N$ thermal baths, the evolution can be represented by 
\begin{equation}~\label{Eq:convex}
\Lambda_n=p_1\Lambda^1_n+p_2\Lambda^2_n+\ldots + p_N\Lambda^N_n
\end{equation}
regarding the convexity of the CP-divisible dynamical maps where $p_i\neq 0$ are the probabilities of experiencing the system from the thermal unit of the $ith$ reservoir. Note that (\ref{Eq:convex}) is the discrete evolution (collision model) equivalent of the continuous Lindblad dynamics expressed in (\ref{Eq:ConvexL}).

\subsection{Physical model}

We also suggest a physical model and discuss its feasibility. The physical model depends on the superconducting circuits and the related architecture~\cite{wendin_quantum_2017}. A two thermal bath example will be analysed by using three transmon qubits including the system qubit. Transmon qubits are the later versions of charge qubits (Cooper pair box) depending Josephson Junction tunnelling devices~\cite{makhlin_quantum-state_2001}. In the physical model, two thermal reservoirs are mimicked by two transmon qubits by repeated interactions process in which they interact through a resonator bus~\cite{koch_charge-insensitive_2007} also serves for the readout~\cite{bianchetti_dynamics_2009}. 

An artificial quantum system  composed of $N$ transmon qubits coupled through a coplanar waveguide (CPW) resonator can be represented by an Hamiltonian
\begin{align}\label{Transmon}
\mathcal{H}=&\omega_r\hat{a}^{\dagger}\hat{a}+\sum_{i=1}^{N} \left[ E_{c_i}(\hat{n}_i-n_{g_i})^2-E_{J_i}\cos \hat{\varphi}_i\right]  +\sum_{i=1}^{N} g_i \hat{n}_i (\hat{a}+\hat{a}^{\dagger}) 
\end{align}
where $\omega_r$ is the resonator frequency,  $\hat{a}$ and $\hat{a}^{\dagger}$ are, respectively, the annihilation and creation operators of the quantum oscillator. The second term of the Hamiltonian corresponds to the transmon qubits where 
where $\hat{n}_i$ is the charge number operator, $n_{g_i}$ is the offset charge and $\hat{\varphi}_i$ the quantized flux of the $ith$ qubit. Here, the flux is defined by $\varphi_i=\pi \Phi_i/\Phi_0$ where $\Phi_i$ is the externally tunable flux of each qubit and $\Phi_0$ is the elementary quanta. Respectively, Josephson energy $E_{J_i}$ and the capacitive energy $E_{c_i}$ are set $E_{J_i}\gg E_{c_i} $ such that the qubits operate in the transmon regime. The last term of the Hamiltonian indicates that the transmon qubits are coupled to the resonator by a strength $g_i$. 

\begin{figure}
\centering
\includegraphics[width=3.0 in]{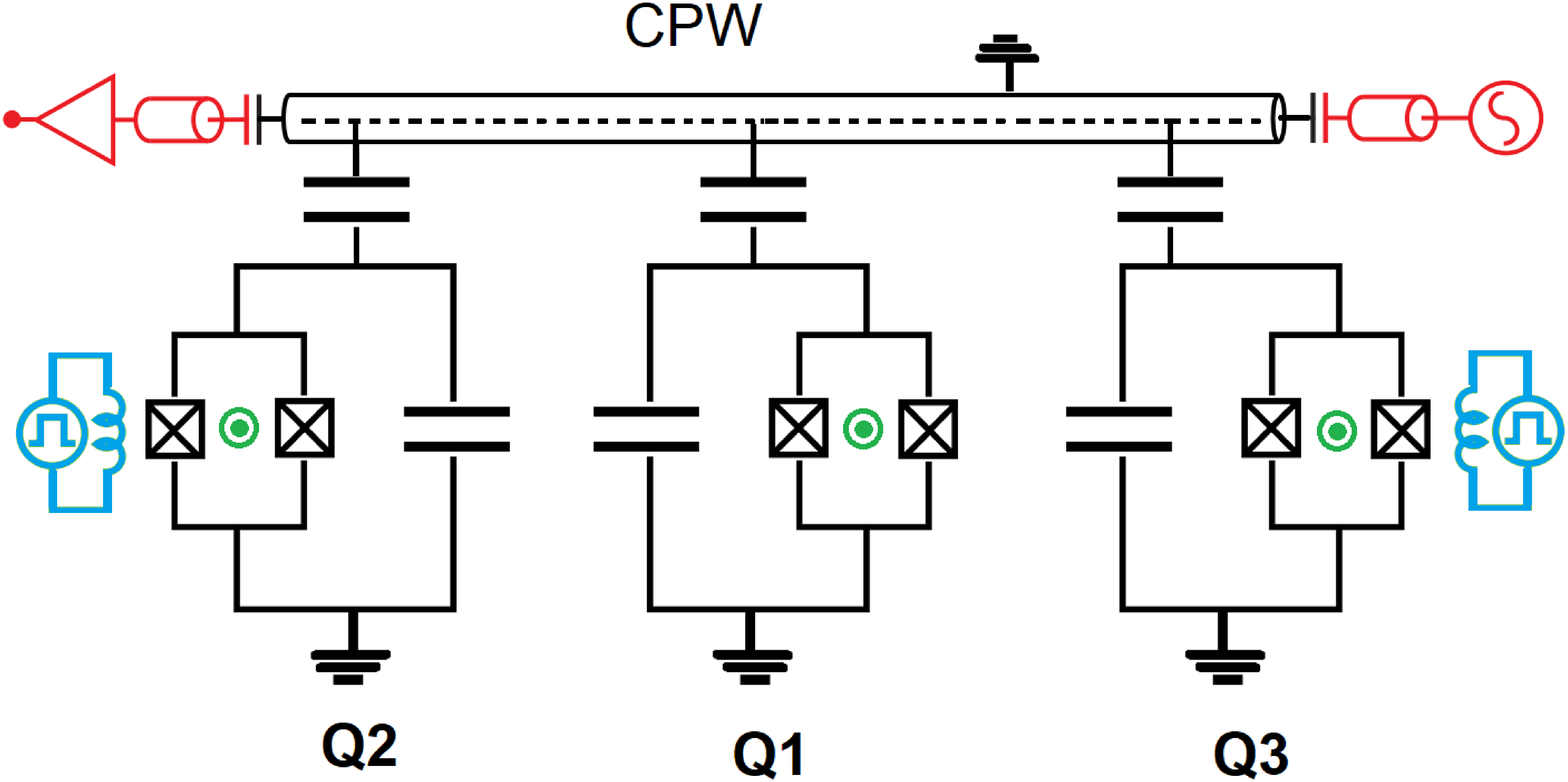} 
\caption{\label{fig:Fig8}(Colour online) Physical model of the classifier through the Lumped-element circuit diagram. Three transmon qubits ($Q_1$ is the system qubit, $Q_2$ and $Q_3$ are the reservoir qubits) are coupled to the superconducting CPW resonator in which they serve for both the readout of the qubits (red) and as a coupling bus. Green dots stand for the flux tunability of each qubit. The tunability allows for the control of coupling to the bus. Microwave lines (blue) are the control fields acting on reservoir qubits $(Q_2, Q_3)$ and they are used for resetting and initialization of ancilla quantum states. } 
\end{figure} 
Here, one can have an issue with the Hamiltonian as the qubit interaction term is not apparent as is in (\ref{Flip}). However, it's known that (\ref{Flip}) type flip-flop interaction could be achieved by the dispersive coupling of the qubits with the same resonator satisfying $\vert \Delta_{1,2,3} \vert=\vert \omega_{1,2,3}-\omega_r \vert\gg g_{1,2,3}$. 

Such a coupling generates an effective interaction between the transmon qubits that they do not directly interact. For instance, the interaction strength between the system qubit (Q1) and one of the reservoir qubits (Q2)  can be described by~\cite{majer_coupling_2007,filipp_multimode_2011}
\begin{equation}\label{Eq:couple}
J_{1,2}=\frac{g_1 g_2}{2} \left( \frac{1}{\Delta_1}+\frac{1}{\Delta_2}\right).
\end{equation}
On the other hand, if the qubit frequencies are made largely dispersive $\vert \omega_1-\omega_2 \vert \gg J_{1,2}$
this effective interaction is effectively turned off. Therefore, inter-qubit couplings can be controlled by qubit frequency settings. Additional requirements should be performed in order to implement the physical model in Fig. 8 successfully. First, on the one hand, the reservoir qubits $Q_2$ and $Q_3$ should interact with the system qubit $Q_1$, on the other hand, $Q_2$ and $Q_3$ should never interact with each other. Moreover, a repetitive switch mechanism should be achieved between the interacting qubits for suitable qubit preparation and reset scenarios. The former could be simply achieved by tuning the ancilla qubit frequencies $\vert \omega_2-\omega_3 \vert\gg J_{2,3}$ largely dispersive. Hence, one obtains 
\begin{align}\label{Eq-effective}
\mathcal{H}=&\frac{\omega_i}{2}\sum_{i=1}^3\sigma_z^i+(\omega_r+\chi_i\sum_{i=1}^3 \sigma_z^i)\hat{a}^{\dagger}\hat{a}+J_{1,i}\sum_{i=2,3}(\sigma_1^+\sigma_i^- +H.c.)
\end{align} 
where $\sigma_z^i$ and $\sigma_i^{\mp}$ are the Pauli operators acting on the subspace belonging to the first two-levels of the respective superconducting qubit and $\chi_i$ are the qubit-dependent resonator frequency shift in which coupled qubit-resonator pairs have no energy exchange between them. 

The switch on/off mechanism can be performed by the externally tunable
magnetic flux $\Phi_i$~\cite{liao_single-particle_2010}. The coupling between the CPW and the qubits can be turned off, by very largely detuning the qubit, using the bias flux and the coupling can be regenerated again by tuning the bias $\Phi_i$  such that the desired dispersive coupling is re-established. There are several timescales for the process of implementing a repeated interaction scheme. The first one is $\tau_{int}$ qubit-CPW interaction time, the second   $\tau_{pr}$ is both the qubit reset and preparation time and $\tau_r$ is the qubit reset time. Therefore the time elapsed between two switch on/off process can be defined as $\mathcal{T}\leq t_{i+1}-t_i$ where $\mathcal{T}=\tau_{int}+\tau_{pr}+\tau_r$. Regarding the reservoir qubits, as the relaxation time $\tau_r$ is much longer than the collision time $\tau_r \gg \tau_{int}$, relaxation of the qubits have no effect between any successive collision times. On the other hand, a strong field can achieve the qubit reset and preparation much shorter than the collision time $\tau_{pr}\ll \tau_{int}$~\cite{liao_single-particle_2010}. Then approximately, one can define the time between any successive switch-on operation as $ t_{i+1}-t_i=\mathcal{T}\simeq\tau_{int}$. 

According to the scenario, the ancilla qubits $Q_2$, $Q_3$ and the system qubit $Q_1$ couple to the CPW (with strength $g$) dispersively. During this coupling process the system generates effective $J_{1,2}$ and $J_{1,3}$ couplings between $Q_1-Q_2$ and $Q_1-Q_3$ in compliance with (\ref{Eq-effective}). The reservoir qubits are initially prepared in their thermal quantum state before the switch-on operation $Q_1$ is prepared in the ground state defines a zero temperature. At time $t_i=0$ the coupling between $Q_1$, $Q_2$, $Q_3$ and CPW turns on and the couplings turns off after $\tau_{int}$. After that, $Q_1$ and $Q_3$ are reset to their initial states. Therefore, system qubit $Q_1$ is now in the tensor product state, that is, $Q_1$ is decoupled from the reservoir qubits and ready for the next collision. Many repetitions of this scheme yield the superconducting circuit model of the thermal quantum classifier composed of transmon qubits. 

As an example, the inverse temperatures of the ancilla qubits representing the thermal reservoirs can be $\beta_{Q_2}=2.898$ ($T_{Q_2}=200$ mK) and $\beta_{Q_3}=4.797$ ($T_{Q_3}=100$ mK)
~\cite{liao_single-particle_2010}. In the weak coupling regime, typically, the resonator frequency is $\omega_r\sim 1-10$ GHz and the coupling between the transmon qubit-resonator is $g\sim 1-500$ MHz~\cite{majer_coupling_2007,wendin_quantum_2017,deng_robustness_2017}. In this frequency regime, a single qubit gate rotation performs within couples of $ns$s. It's been shown that 2000 collisions, by two reservoir qubits as in Fig. 8, can equilibrate a single spin taking each collision time $\tau=5 ns$~\cite{turkpence_steady_2019}. This result corresponds to $\sim 7.5-10\mu$s to reach the steady state. 

Current state-of-the-art allows for $T_1\sim 20-60$ $\mu$s energy relaxation time for transmon qubits~\cite{deng_robustness_2017,kirchhoff_optimized_2018}. That is, the physical model of the thermal quantum classifier proposed in this section can reliably classify the temperature information before the information is lost. As a result of these considerations, we underline that the response time of the physical classifier is in the $\mu$s range. Comparing with the speed of the classical classifiers in which operate by a $m$s CPU time~\cite{Enrico_2009}, one concludes that the physical model of the quantum classifier operates three orders of magnitude faster than the classical ones.

\section{Conclusions}

We have proposed and numerically demonstrated an open quantum model classifies the temperature information of the thermal environments in the weak coupling regime. The theoretical model relies on the additivity of the quantum dynamical maps while physical model relies on the divisibility of the quantum maps. We also analytically obtained that a single qubit system, coupled to several thermal environments with finite temperatures, returns an arithmetic mean temperature in the steady state in the case of equal coupling rates. Moreover, we numerically demonstrated that, depending on the modification of the coupling rates, the system response temperature could be larger or smaller than the arithmetic mean of the thermal environment temperatures. That is, we showed that the proposed single qubit quantum model is a binary classifier. 

We also report that the obtained response character of the proposed quantum classifier fit with the linear activation functions. We choose the superconducting circuits and the repeated interactions scheme, respectively, for the physical model and the representation of the open quantum dynamics. Three transmon qubit example, with one system qubit and two environment qubits in the microwave regime, was given as a physical model example. The capability of the temperature data response depending on the environment temperatures could improve the studies about building the smallest thermal machines as well as temperature sensing in these small scales. 

\section{Acknowledgements}
Authors acknowledge support from \.{I}stanbul Technical University. This research did not receive any specific grant from funding agencies in the public, commercial, or not-for-profit sectors.

\appendix
\section{The steady state temperature}

Here, we present a derivation of the steady state temperature of the single spin in contact with $N$ thermal reservoirs. To this end, we solve (\ref{MasterEq}) to obtain the response of the two-level system in the form of (\ref{QbitTemp}). First, we write (\ref{MasterEq}) for $N=2$ thermal reservoirs. The master equation reads
\begin{align}\label{APPEq1}
 \dot{\varrho}=&-i[\frac{\omega}{2}\sigma_z,\varrho]+\frac{\Gamma^{(1)}}{2}(\bar{n}^{(1)}+1)(2\sigma^{-}\varrho\sigma^{+}-\sigma^{+}\sigma^{-}\varrho\nonumber -\varrho\sigma^{+}\sigma^{-})\nonumber 
 \\&+\frac{\Gamma^{(1)}}{2}\bar{n}^{(1)}(2\sigma^{+}\varrho\sigma^{-}-\sigma^{-}\sigma^{+}\varrho
-\varrho\sigma^{-}\sigma^{+})\nonumber\\
&+\frac{\Gamma^{(2)}}{2}(\bar{n}^{(2)}+1)(2\sigma^{-}\varrho\sigma^{+}-\sigma^{+}\sigma^{-}\varrho-\varrho\sigma^{+}\sigma^{-})\nonumber\\
&+\frac{\Gamma^{(2)}}{2}\bar{n}^{(2)}(2\sigma^{+}\varrho\sigma^{-}-\sigma^{-}\sigma^{+}\varrho-\varrho\sigma^{-}\sigma^{+})
\end{align}
where we take $\hbar=1$. We remind that we define the Pauli operators as $\sigma^{-}=\ket{g}\bra{e}$, $\sigma^{+}=\ket{e}\bra{g} $ and $\sigma_z=\ket{e}\bra{e}-\ket{g}\bra{g}$. Next, we take the commutation, the first term of (\ref{APPEq1}) and take $\dot{\varrho}=0$ since we seek the solution in the steady state. By these specifications and regarding $\langle \nu_i | \nu_j\rangle=\delta_{ij}$ where $\ket{\nu_{i,j}}$ are the orthogonal basis states, we have

\begin{align}\label{APPEq2}
 \dot{\varrho}=&\frac{-i\omega}{2}\left(\ket{e}\bra{e}\varrho-\ket{g}\bra{g}\varrho-\varrho\ket{e}\bra{e}+\varrho\ket{g}\bra{g}\right)\nonumber\\
&+\frac{\Gamma^{(1)}}{2}\bar{n}^{(1)}(2\ket{g}\bra{e}\varrho\ket{e}\bra{g}-\ket{e}\bra{e}\varrho- \varrho\ket{e}\bra{e}) \nonumber \\
&+\frac{\Gamma^{(1)}}{2}(2\ket{g}\bra{e}\varrho\ket{e}\bra{g}-\ket{e}\bra{e}\varrho- \varrho\ket{e}\bra{e})\nonumber\\ 
&+\frac{\Gamma^{(1)}}{2}\bar{n}^{(1)}(2\ket{e}\bra{g}\varrho\ket{g}\bra{e}-\ket{g}\bra{g}\varrho-\varrho\ket{g}\bra{g}) \nonumber \\
&+\frac{\Gamma^{(2)}}{2}\bar{n}^{(2)}(2\ket{g}\bra{e}\varrho\ket{e}\bra{g}-\ket{e}\bra{e}\varrho-\varrho\ket{e}\bra{e}) \nonumber\\
&+\frac{\Gamma^{(2)}}{2}(2\ket{g}\bra{e}\varrho\ket{e}\bra{g}-\ket{e}\bra{e}\varrho-\varrho\ket{e}\bra{e}) \nonumber \\
&+\frac{\Gamma^{(2)}}{2}\bar{n}^{(2)}(2\ket{e}\bra{g}\varrho\ket{g}\bra{e}-\ket{g}\bra{g}\varrho-\varrho\ket{g}\bra{g})=0.
\end{align}
We define the matrix elements $\bra{e}\varrho\ket{e}=p_e$ and $\bra{g}\varrho\ket{g}=p_g$, respectively, as the excited and the ground state populations. Multiplying the left-hand side of (\ref{APPEq2}) by $\bra{e}$ or $\bra{g}$ and then the right-hand side by $\ket{e}$ or $\ket{g}$, we have 
\begin{equation}\label{APPEq3}
\frac{p_g}{p_e}=\frac{\bar{n}^{(1)}\Gamma^{(1)}+\bar{n}^{(2)}\Gamma^{(2)}+\Gamma^{(1)}+\Gamma^{(2)}}{\bar{n}^{(1)}\Gamma^{(1)}+\bar{n}^{(2)}\Gamma^{(2)}}.
\end{equation}
Note that, this is only the $N=2$ two reservoirs expansion of (\ref{MasterEq}). It's straightforward that in the general case for $N$ thermal reservoirs we reach (\ref{Ratio}). We continue with the $N=2$ reservoir case where the couplings are $\Gamma^{(1)}=\Gamma^{(2)}$ equal. In that case, (\ref{APPEq3}) becomes
\begin{equation}\label{APPEq4}
\frac{p_g}{p_e}=\frac{\sum\bar{n}+2}{\sum\bar{n}}
\end{equation}
where ${\sum \bar{n}}=\bar{n}^{(1)}+\bar{n}^{(2)}$. We take the natural logarithm both sides of (\ref{APPEq4}) as
\begin{equation}\label{APPEq5}
ln\left(\frac{p_g}{p_e}\right)=ln\left(\sum\bar{n}+2\right)-ln\left(\sum\bar{n}\right)
\end{equation}
For simplicity, denoting $\sum \bar{n}=\mathcal{N}$, we rewrite the first term of equation (\ref{APPEq5}) as 
\begin{equation}\label{APPEq6}
ln\left(\mathcal{N}+2\right)=ln\left(\mathcal{N}(1+\frac{2}{\mathcal{N}})\right)\approx ln\left(\mathcal{N}\right)+\frac{2}{\mathcal{N}}
\end{equation}
where we have used a simple logarithm identity and an approximation such as, respectively, $log(AB)=log(A)+log(B)$ and $ln(1+x)\approx x$. Inserting this result into equation (\ref{APPEq5}) we have 
\begin{equation}\label{APPEq7}
ln\left(\frac{p_g}{p_e}\right)=\frac{2}{\mathcal{N}}.
\end{equation}
Next, we expand this result as $\mathcal{N}=\bar{n}^{(1)}+\bar{n}^{(2)}$ where $\bar{n}^{(i)}=1/(\texttt{exp}(\omega/T^{(i)})-1)$ as we defined in equation (\ref{Excitation}). Using the high temperature approximation $\texttt"{exp}(\omega/T^{(i)})\approx \omega/T^{(i)}+1$, equation (\ref{APPEq7}) becomes

\begin{equation}
ln\left(\frac{p_g}{p_e}\right)\approx \frac{2}{\frac{T^{(1)}}{\omega}+\frac{T^{(2)}}{\omega}}=\frac{2\omega}{T^{(1)}+T^{(2)}}.
\end{equation}
Inserting this result into the definition of qubit temperature in equation (\ref{QbitTemp}), finally, we have the analytical expression of the steady temperature response of the proposed quantum classifier for $N=2$ thermal reservoirs.

\begin{equation}
(T_S)^{ss}\cong \frac{T^{(1)}+T^{(2)}}{2}.
\end{equation}
Note that, again, it's straightforward to generalize this result for an arbitrary number of thermal reservoirs just as in (\ref{AvarageT}).




\end{document}